\documentclass[iop]{emulateapj}
\usepackage{apjfonts}
\usepackage{amsmath}
\usepackage{graphicx}
\usepackage{natbib}
\shorttitle{Understanding Compact Object formation. III.}
\shortauthors{T.-W. Wong et al.}

\begin{document}

\title{Understanding Compact Object Formation and Natal Kicks. III. The case of Cygnus X-1}

\author{Tsing-Wai Wong$^1$, Francesca Valsecchi$^1$, Tassos Fragos$^2$, Vassiliki Kalogera$^1$}

\affil{
$^1$Center for Interdisciplinary Exploration and Research in Astrophysics (CIERA) \& Department of Physics and Astronomy, Northwestern University,\\
2145 Sheridan Road, Evanston, IL 60208, USA; tsingwong2012@u.northwestern.edu, francesca@u.northwestern.edu,vicky@northwestern.edu\\
$^2$Harvard-Smithsonian Center for Astrophysics, 60 Garden St., Cambridge, MA 02138, USA; tfragos@cfa.harvard.edu}

\begin{abstract}
In recent years, accurate observational constraints become available for an increasing number of Galactic X-ray binaries. Together with proper motion measurements, we could reconstruct the full evolutionary history of X-ray binaries back to the time of compact object formation. In this paper, we present the first study of the persistent X-ray source Cygnus\;X-1 that takes into account of all available observational constraints. Our analysis accounts for three evolutionary phases: orbital evolution and motion through the Galactic potential after the formation of black hole (BH), and binary orbital dynamics at the time of core collapse. We find that the mass of the BH immediate progenitor is $15.0 - 20.0$ M$_\sun$, and at the time of core collapse, the BH has potentially received a small kick velocity of $\le 77$ km s$^{-1}$ at 95\% confidence. If the BH progenitor mass is less than $\sim 17$ M$_\sun$, a non zero natal kick velocity is required to explain the currently observed properties of Cygnus\;X-1. Since the BH has only accreted mass from its companion's stellar wind, the negligible amount of accreted mass is impossible to explain the observationally inferred BH spin of $a_* > 0.95$, and the origin of this extreme BH spin must be connected to the BH formation itself. Right after the BH formation, we find that the BH companion is a $19.8 - 22.6$ M$_\sun$ main sequence star, orbiting the BH at a period of $4.7 - 5.2$ days. Furthermore, recent observations show that the BH companion is currently super-synchronized. This super-synchronism indicates that the strength of tides exerted on the BH companion should be weaker by a factor of at least two compared to the usually adopted strength.
\end{abstract}


\section{Introduction}

In recent years, the number of observed black hole (BH) X-ray binaries (XRBs) has grown significantly. For these binaries, there exists a wealth of observation information about their current physical state:  BH and donor masses, orbital period, donor's position on the H-R diagram and surface chemical composition, transient or persistent X-ray emission, and Roche lobe overflow (RLO) or wind-driven character of the mass transfer (MT) process. Furthermore, proper motions have been measured for a handful of these binaries \citep[e.g.][]{Mirabel...01, Mirabel...02, Mirabel-Rodrigues03}. Together with the earlier measurements of center-of-mass radial velocities and distances, we can obtain information about the three-dimensional kinematic properties of these binaries. Given this plethora of observation results, the current observed sample of BH XRBs provides us with a unique opportunity to understand the formation and evolution of BHs in binaries. This paper is the third in a series where we investigate in detail the BH formation in XRBs, especially focusing on the mass relationship between BHs and their immediate progenitors and the possible BH natal kick magnitude imparted during the core collapse event.

In the first paper of this series, \citet[][hereafter Paper I]{Bart...05} showed how using the currently available constraints one could uncover the evolution history of an XRB from the present state back to the time just prior to the core collapse event. They applied their analysis to the BH XRB GRO J1655-40. In the second paper, \citet[][hereafter Paper II]{Tassos...09} performed the same analysis for the case of the BH XRB XTE J1118+480. In this work, we focus on the case of the BH XRB Cygnus X-1. The mass transfer mechanism in Cygnus X-1 is different from the XRBs studied in previous cases. Both donors in GRO J1655-40 and XTE J1118+480 are transferring mass to the BH under Roche lobe overflow, whereas the BH in Cygnus X-1 is accreting mass from the stellar wind of its companion.

The plan of the paper is as follows. In Section 2, we review Cygnus X-1's currently available observational constraints. A general outline of the analysis used to reconstruct the system's evolutionary history is presented in Section 3, while individual steps of the analysis are discussed in more detail in Section $4 - 7$. In Section 8, we derive constraints on the formation of the BH. The final section is devoted to a summary of our results and discussion of some of the assumptions introduced in our analysis.

\section{Observational Constraints for Cygnus\,X-1}

Cygnus X-1 was first detected in Aerobee surveys in 1964 by  \cite{Bowyer...65}. Soon after the discovery, it was identified as an XRB, which consisted of a compact object and a visible star HDE 226868 \citep{murdin-webster71, webster-murdin72, bolton72a}. Spectroscopic observations led \cite{walborn73} to classify HDE 226868 as an O9.7 Iab supergiant. \cite{bregman...73} estimated the distance to be 2.5 kpc and set a lower limit of 1 kpc, based on the colors of field stars in the vicinity of the supergiant. Using a combination of data from David Dunlap Observatory (DDO) and the Royal greenwich Observatory, \cite{bolton72b} derived the orbital period, eccentricity and systemic radial velocity (V$_0$) to be $5.5995 \pm{0.0009}$ days, $0.09 \pm{0.02}$ and $-6.0 \pm{0.1}$ km s$^{-1}$, respectively. 
Based on the absence of X-ray and optical eclipses, the author gave a lower limit of 7.4 M$_\sun$ on the mass of the compact object. This implied the compact object was too massive to be a white dwarf or a neutron star. Thus, the author confirmed \citet{webster-murdin72}'s finding that the compact object is a very strong BH candidate.

Using the orbital period obtained from spectrometry and a range in the assumed degree of Roche filling of the supergiant, \cite{gies-bolton82, gies-bolton86} found a lower mass limit of 7 M$_\sun$ for the compact object. This confirmed that the compact object observed in Cygnus X-1 was a BH. The same authors also refined the orbital period and eccentricity to be $5.59974 \pm{0.00008}$ and $0.021 \pm{0.013}$, respectively, and measured V$_0$ to be $-2.0 \pm{0.7}$ km s$^{-1}$. \cite{ninkov...87} used the relationship between the equivalent width of the H$\gamma$ spectral line and the absolute magnitude of early-type supergiants to estimate the distance as $2.5 \pm{0.3}$ kpc .

\cite{herrero...95} performed a detailed spectroscopic analysis on the supergiant, and derived the masses to be 10.1 M$_\sun$ and 17.8 M$_\sun$ for the BH (M$_{\rm BH}$) and the supergiant (M$_2$), respectively, if an orbital inclination angle of 35$^\circ$ was assumed. Using the Isaac Newton telescope, \cite{lasala...98} measured the orbital period as $5.5997 \pm{0.0001}$ days and V$_0$ as $-5.4 \pm{0.1}$ km s$^{-1}$. With all the accumulated radial velocity measurements and their own spectroscopy of the supergiant, \cite{brocksopp...99} refined the orbital period to $5.599829 \pm{0.000024}$. The proper motion of Cygnus X-1 was observed with the Very Large Baseline Interferometry (VLBI) between 1988 and 2001 \citep{lestrade...99, stirling...01, Mirabel-Rodrigues03}. During this period, the system's position shifted at a rate of $-4.2 \pm{0.2}$ mas yr$^{-1}$ in right ascension (R.A.) and $-7.6 \pm{0.2}$ mas yr$^{-1}$ in declination (dec.). Meanwhile, a trigonometric parallax of $0.73 \pm{0.30}$ mas was also measured with VLBI, which gave a distance of 1.4$^{+0.9}_{-0.4}$ kpc  \citep{lestrade...99}.

By studying the spectra obtained with the 0.9 m coud\'e feed telescope of Kitt Peak National Observatory, the 2.1 m telescope of University of Texas McDonald Observatory, and the 1.9m telescope of University of Toronto David Dunlap Observatory between 1998 and 2002, \cite{gies...03} derived V$_0$ as $-7.0 \pm{0.5}$ km s$^{-1}$ and estimated M$_{\rm BH}$/M$_2$ $\approx 0.36 \pm{0.05}$. \cite{caballero...09} examined the supergiant's ultraviolet spectra from the Hubble space telescope. Their results gave masses of 23$^{+8}_{-6}$ and 11$^{+5}_{-3}$ M$_\sun$ for the supergiant and the BH, respectively. On the other hand, \cite{shaposhnikov-titarchuk07} used the X-ray quasi-periodic oscillation and spectral index relationship and deduced M$_{\rm BH}$ to be $8.7 \pm{0.8}$ M$_\sun$, which overlapped with the lower end of the M$_{\rm BH}$ range derived by \cite{caballero...09}.

Recently, \cite{reid...11} measured the trigonometric parallax of Cygnus X-1 with the National Radio Astronomy Observatory's Very Long Baseline Array (VLBA) and found a distance of 1.86$^{+0.12}_{-0.11}$ kpc. The authors also reported proper motion measurements of Cygnus X-1, which were $-3.78 \pm{0.06}$ mas yr$^{-1}$ in R.A. and $-6.40 \pm{0.12}$ mas yr$^{-1}$ in dec. Meanwhile, \cite{xiang...11} studied the X-ray dust scattering halo of Cygnus\;X-1 and determined the distance to be $1.81 \pm{0.09}$ kpc, after considering the compatibility with the parallax result. Building on the trigonometric parallax distance measurement of \cite{reid...11}, \cite{orosz...11} performed optical data modeling of Cygnus\;X-1, and found the mass of the supergiant to be $19.2 \pm{1.9}$ M$_\sun$ and the black hole to be $14.8 \pm{1.0}$ M$_\sun$. Using the results of \cite{reid...11} and \cite{orosz...11}, \cite{gou...11} determined that Cygnus\;X-1 hosts a near-extreme Kerr BH, with a spin parameter $a_* > 0.95$.

Unlike most of the XRBs known to host a BH, Cygnus X-1 is a persistent X-ray source. Since the supergiant is currently not overfilling its Roche lobe \citep{gies-bolton86}, the observed X-rays are mainly powered by the accretion of stellar wind. The X-ray luminosity of Cygnus X-1 varies between two discrete levels, namely the "hard (low) state" and the "soft (high) state". As the system spends most of its time \citep[$\sim$90\%, see][]{cadolle...06} in the hard state, we focus on the hard state X-ray luminosity (L$\rm _X$). \cite{frontera...01} observed Cygnus X-1 with the Narrow Field Instruments of the BeppoSAX satellite at different epochs in 1996. The authors obtained the $\rm L_X$ ($0.5 - 200$ keV) and the extrapolated bolometric luminosity (L$\rm_{bol}$) as $2.0 \times 10^{37}$ and $2.4 \times 10^{37}$ erg s$^{-1}$, respectively, assuming a distance of 2 kpc. Using observational data obtained by the Compton Gamma Ray Observatory (CGRO) between 1991 and 2000, \cite{mcconnell...02} derived L$\rm_{bol}$ to be (1.62 -- 1.70) $\times$ 10$^{37}$ erg s$^{-1}$, with the distance to the source fixed at 2 kpc. \cite{cadolle...06} observed Cygnus X-1 with the International Gamma-Ray Astrophysics Laboratory (INTEGRAL) between 2002 and 2004 and measured L$\rm _X$ ($20 - 100$ keV) as $6.5 \times 10^{36}$ erg s$^{-1}$, assuming a distance of 2.4 kpc. The authors also gave L$\rm_{bol}$ as $2.2 \times 10^{37}$ erg s$^{-1}$. 

For the systemic parameters relevant to our analysis, we adopt the most recent observational constraints, with the exception of L$\rm_{bol}$. We consider all the L$\rm_{bol}$ values mentioned above, assuming they represent the typical X-ray variability range for this system. After rescaling their values to the parallax distance measurement by \cite{reid...11} and considering the uncertainty in that distance, we adopt L$\rm_{bol}$ to be $(1.17 - 2.35) \times 10^{37}$ erg s$^{-1}$. For ease of reference, our adopted observational constraints are summarized in Table\;1.

\begin{deluxetable*}{lccc}
\tablewidth{18.0 cm}
\tablecolumns{4}
\tablecaption{Properties of Cygnus X-1}
\tablehead{
\colhead{Parameter} & \colhead{Notation} & \colhead{Value} & \colhead{References}
}
\startdata
Distance (kpc) & d & 1.86$^{+0.12}_{-0.11}$ & (9)\\
Galactic longitude (deg) & l & 71.3 & (2)\\
Galactic latitude (deg) & b & +3.1 & (2)\\
Proper motion in R.A. (mas yr$^{-1}$) & $\mu_{\rm R.A.}$ & $-3.78 \pm{0.06}$ & (9)\\
Proper motion in decl. (mas yr$^{-1}$) & $\mu_{\rm decl.}$ & $-6.40 \pm{0.12}$ & (9)\\
Systemic velocity (km s$^{-1}$) & V$_0$ & $-7.0 \pm{0.5}$ & (5)\\
Orbital period (days) & P$_{\rm orb}$ & $5.599829 \pm{0.000016}$ & (1)\\
Orbital eccentricity & e$_{\rm orb}$ & $0.018 \pm{0.003}$ & (8)\\
Inclination angle & $i$ & $27.06 \pm{0.76}$ & (8)\\
Black hole mass (M$_\sun$) & M$_{\rm BH}$ & $14.81 \pm{0.98}$ & (8)\\
Black hole spin & $a_*$ & $> 0.95$ & (10)\\
Companion mass (M$_\sun$) & M$_2$ & $19.16 \pm{1.90}$ & (8)\\
Companion Radius (R$_\sun$) & R$_2$ & $16.50 \pm{0.84}$ & (8) \\
Companion Luminosity (L$_\sun$) & L$_2$ & (1.91 -- 2.75) $\times$ 10$^5$ & (8)\\
Companion Effective temperature (K) & T$_{\rm eff}$ & 30000 -- 32000 & (8)\\
Companion surface rotation speed (km s$^{-1}$) & V$_{\rm rot} \sin i$ & $95 \pm{6}$ & (7)\\
Bolometric luminosity of the X-ray source (erg s$^{-1}$) & L$\rm_{bol}$ & $\rm (1.3 - 2.1) \left (\frac{d}{1.86 \; kpc} \right )^2 \times 10^{37}$ & (3),(4),(6)\\
\enddata

\tablerefs{(1) Brocksopp et al. 1999, (2) Lestrade et al. 1999, (3) Frontera et al. 2001, (4) McConnell et al. 2002, (5) Gies et al. 2003, (6) Cadolle Bel et al. 2006, (7) Caballero-Nieves et al. 2009, 
(8) Orosz et al. 2011, (9) Reid et al. 2011, (10) Gou et al. 2011} 
\end{deluxetable*}

\section{Outline of Analysis Methodology}

In our analysis, we assume that Cygnus X-1 formed in the Galactic disk, from the evolution of an isolated primordial binary at solar metallicity. In fact, \cite{Mirabel-Rodrigues03} suggest that Cygnus X-1 belongs to Cygnus OB3 (Cyg OB3), which is an OB association located close to the Galactic plane. We also assume that no mass transfer via Roche lobe overflow occurred in the evolutionary history of this binary.

According to our current understanding, in order to form a $\sim 15$ M$_\sun$ stellar BH at solar metalicity, the BH progenitor in the primordial binary needs to be more massive than 120 M$_\sun$ \citep{belczynski...10}. Such a massive star loses its hydrogen rich envelope via stellar wind, and exposes its naked helium core. At the end of nuclear evolution, it collapses into a BH. During the core collapse event, the orbit is altered by the asymmetric mass loss from the system and a possible recoil kick imparted to the BH. If the binary survives through the core collapse event, angular momentum loss via gravitational radiation and tidal effects causes the orbit to shrink, although wind mass loss leads to orbital expansion. In the meantime, the more evolved BH companion is losing mass via its own stellar wind at a higher rate. The system becomes a BH XRB when the BH captures a non-negligible amount of mass from its companion's stellar wind. 

In this paper, we restrict ourselves to the formation of BH XRBs through the above evolutionary channel. Like the first two papers, our goal is to track the evolutionary history of Cygnus X-1 back to the time just prior to the core collapse event. Our analysis incorporates a number of calculations which can be summarized in four steps.

First, we identify the current evolutionary stage of the BH companion, so that all the observational constraints are satisfied. Under the assumption that the BH companion mass has not been altered by mass transfer in the past, we model it as an isolated star. Using a stellar evolution code, we calculate a grid of evolutionary sequences of isolated stars at different zero age main sequence (ZAMS) masses. We examine each sequence to find whether there exists a point in time that the calculated stellar properties, i.e. mass, radius, luminosity and effective temperature, are all simultaneously in agreement with the currently observed properties of the BH companion. If such a period of time exists, we classify that sequence as "successful". The current age of the BH companion can be estimated from these successful sequences, and the time expired since the BH formation can then be derived by subtracting the approximate lifetime of the BH progenitor. 

Next, we consider the kinematic evolutionary history of the XRB in the Galactic potential. Starting from the current location, we follow the methodology of \cite{gualandris...05} and use the observed three-dimensional velocity to trace the Galactic motion of Cygnus X-1 backward in time. Together with the constraints on the current age of the system derived in the first step, this allows us to determine the location and velocity of the binary at the time of BH formation (we denote these as "birth" location and velocity). By subtracting the local Galactic rotational velocity at the "birth" location from the systems's center-of-mass velocity, we derive constraints on the \em peculiar \em velocity of the binary right after the formation of the BH. 

In the third step, we analyze the orbital dynamics of the core collapse event due to mass loss and possible natal kicks imparted to the BH. In this paper, we refer to the instants right before and after the formation of the BH by the terms "pre-SN" and "post-SN", respectively. We start with the constrained parameter space of (M$_{\rm BH}$, M$_2$) derived in the first step and perform a Monte Carlo simulation scanning over the parameter space of the pre-SN binary properties. This parameter space is limited by requirements of orbital angular momentum and energy conservations, and by the post-SN binary peculiar velocity constraint derived in the second step. This calculation yields a population of simulated post-SN binaries for each successful sequence.

Finally, we follow the orbital evolution of these simulated binaries to the current epoch. Our calculation accounts for tides, wind mass loss, wind accretion onto the BH, and orbital angular momentum loss via gravitational radiation. At the end of the calculations, we require agreement between the observed and calculated orbital period and eccentricity.

\section{Modeling the BH Companion}

Under the assumption that the companion mass has not been altered by mass transfer in its past, we model the companion as an isolated star using a modified version of the stellar evolution code \textit{EZ} \cite[originally developed by][]{paxton04}.

We calculate the evolution of our stellar models at solar metallicity, which is the same metallicity that \cite{orosz...11} used in deriving the properties of the BH companion. When we place the companion's observational constraints on an H-R diagram, we find that the current location of the companion does not seem to be consistent with any evolutionary tracks calculated by the stellar evolution code. As shown in Figure\;1, the companion is overluminous for a star of its mass. This cannot be explained by earlier mass transfer from the BH progenitor to the companion. \cite{braun-langer95} studied the effects of mass accretion onto massive main sequence stars, and found that the accreting stars would not appear overluminous for their new masses during the rest of their main sequence lifetime. If mass accretion leads to a so called "rejuvenation" of the accreting star, which means its central hydrogen abundance substantially increases, it would have the same luminosity as a star of its new mass. If rejuvenation does not occur, the accreting star would appear underluminous for its new mass during the rest of its main sequence lifetime.

\begin{figure}
\begin{center}
\plotone{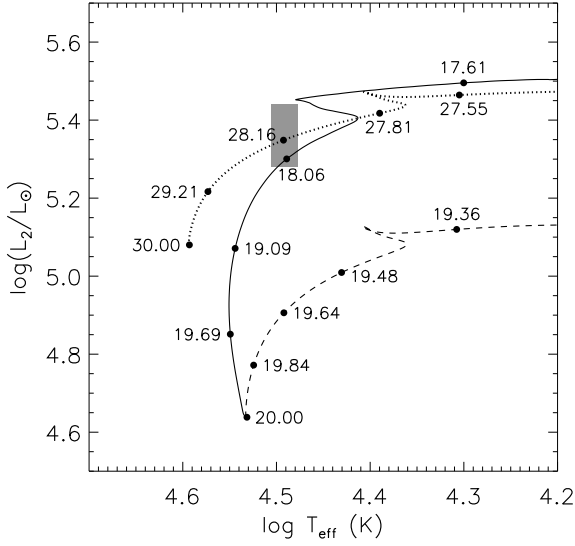}
\end{center}
\caption{The evolutionary tracks for isolated stars on the H-R diagram. On each track, the mass of the star in M$_\sun$ is indicated at various points. The gray shaped area represents the observational constraints of the BH companion. At T$\rm_{eff} \approx 31000$K, the model with an initial mass of 20 M$_\sun$ (dashed line) would have a mass $\approx 19.64$ M$_\sun$, which is in good agreement with the measured mass of the companion. However, it does not match the measured luminosity. On the other hand, the model with an initial mass of 30 M$_\sun$ (dotted line) match the measured luminosity at T$\rm_{eff} \approx 31000$ K, but the mass of the star does not match the companion's. The model with an initial mass of 20 M$_\sun$ and $\rm \alpha_{ov} = 0.45$ (solid line) could match both the measured mass and luminosity of the companion at T$\rm _{eff} \approx 31000$ K.}
\end{figure}

One possible solution for matching the observed companion's luminosity is increasing the core overshooting parameter $\alpha_{\rm ov}$ to $\sim 0.45$. Although this value is relatively high, it is not unphysical. \cite{claret07} compared the data from 13 double-line eclipsing binary systems with theoretical predictions of stellar modeling and found $\alpha_{\rm ov}$ could be as high as 0.6 for massive stars. We vary $\alpha_{\rm ov}$ from 0.35 to 0.5, in steps of 0.01. We note that the need for such higher values of $\alpha_{\rm ov}$ in the modeling of massive stars may very well be connected to the significant presence of internal rotation and associated rotational mixing. Effectively increasing $\alpha_{\rm ov}$ leads to stronger internal mixing and in a way allows the stellar model to behave more like a rotating model.

Besides the observational constraints on the companion's properties, there are three additional constraints. The first one comes from the fact that the companion is currently not overfilling its Roche lobe \citep{gies-bolton86}. Thus, we require the stellar radius R$_2$ in our models to be
\begin{equation}
R_2 \; \leq \; A_{\rm orb}r_{\rm Egg} + \Delta R \ ,
\end{equation}
where r$_{\rm Egg}$ is the effective Roche lobe radius given by \cite{eggleton83}. Here, we make an approximation that the orbit is circular and synchronized. The parameter $\Delta$R is a constant accounting for the difference in the calculated stellar radii among stellar evolution codes \citep{valsecchi...10}. We set $\Delta$R to 2.5 R$_\sun$.

Another constraint is that the calculated bolometric luminosity (L$_{\rm bol}$) resulting from the stellar wind accretion process needs to fall within the observational range, which is $(1.17 - 2.35) \times 10^{37}$ erg s$^{-1}$. By adopting the \cite{bondi-hoyle44} accretion model and following \cite{belczynski...08}, the orbital-averaged accretion rate is given by
\begin{equation}
\dot{M}_{acc} = - \frac{F_{wind}}{\sqrt{1-e_{orb}^2}}  \left( \frac{GM_{BH}}{V_{wind}^2} \right)^2 \frac{\alpha_{wind}}{2A_{orb}^2} \frac{\dot{M}_2}{(1+V^2)^{3/2}}
\end{equation}
Here, M$_{\rm BH}$ is the BH mass in our models, which varies within the $1\sigma$ range of the observational constraint, in steps of 0.098 M$_\sun$. Since the total mass that the BH could have accreted from its companion stellar wind is negligible, M$\rm_{BH}$ in each evolutionary sequence is fixed throughout our analysis. $\rm \dot{M}_2$ is the wind mass loss rate of the companion in our models. F$_{\rm wind}$ is a parameter such that $\rm \dot{M}_{acc}$ never exceeds $-0.8 \rm \dot{M}_2$, and $\alpha_{\rm wind}$ is the accretion efficiency, which varies between 1.5 and 2.0 \citep{boffin-jorissen88}. A$_{\rm orb}$ and e$_{\rm orb}$ are the orbital semi-major axis and eccentricity, respectively. A$_{\rm orb}$ is derived from the mean measured orbital period $\rm P_{orb}$, which is
\begin{equation}
A_{orb} = \left [ {\frac{G \left ( M_{BH} + M_2 \right ) P_{orb}^2}{4\pi^2}} \right ] ^{\frac{1}{3}},
\end{equation}
where M$_2$ is the companion mass in our models. e$\rm_{orb}$ is set equal to the mean measured orbital eccentricity. V$_{\rm wind}$ denotes the wind velocity. V$^2$ equals to V$_{\rm BH}^2$/V$_{\rm wind}^2$, where V$_{\rm BH}^2$ is the orbital velocity square of the BH and is approximated as G(M$_{\rm BH}$ + M$_2$)/A$_{\rm orb}$. We adopt the spherically symmetric wind velocity law given in \cite{lamers-cassinelli99},
\begin{equation}
V_{\rm wind}(r) \ = \ V_{\rm esc} \ + \ (V_\infty \ -\  V_{\rm esc})\left(1 \ - \ \frac{R_2}{r} \right)^\beta
\end{equation}
where r is the distance from the companion to the BH and is set equal to A$_{\rm orb}$. $\beta$ is a free parameter varying from 0.6 to 1.6 \citep{gies-bolton86b, lamers-leitherer93}, in steps of 0.1. V$_\infty$ is the wind velocity at infinity, while V$_{\rm esc}$ is the effective escape velocity at the surface of the companion. Within the typical range of O star surface temperature, V$_\infty$ is scaled as 2.65V$_{\rm esc}$ \citep{kudritzki-puls00}. Following \cite{lamers-cassinelli99},
\begin{equation}
V_{\rm esc} = \sqrt{2(1-\Gamma_e)GM_2/R_2},
\end{equation}
where
\begin{equation}
\Gamma_e \ = \ \frac{\sigma_eL_2}{4\pi cGM_2}
\end{equation}
is the mass correcting factor for the radiative force due to electron scattering, and c is the speed of light in vacuum. \cite{lamers-leitherer93} scaled the electron scattering coefficient per unit mass $\sigma_e$ as
\begin{equation}
\sigma_e \ = \ 0.401\left (\frac{1\ +\ q\epsilon}{1\ + \ 3\epsilon} \right),
\end{equation}
where q is the fraction of He$^{++}$ and (1 -- q) is the fraction of He$^+$, with q = 1 if T$_{\rm eff}$ $\geq$ 35,000 K, q = 0.5 if 30,000 K $\leq$ T$_{\rm eff}$ < 35,000 K, and q = 0 if T$_{\rm eff}$ < 30,000 K. The abundance ratio $\epsilon$ = He/(H + He) is fixed at 0.15, which is appropriate for an O star with a spectral type of Class I. Using $\rm \dot{M}_{\rm acc}$ from equation (2),  we follow \cite{belczynski...08} and calculate the bolometric luminosity resulting from the companion's stellar wind being accreted onto the BH as
\begin{equation}
L_{\rm bol} = \frac{1}{2} \frac{GM_{\rm BH}\dot{M}_{\rm acc}}{R_{acc}},
\end{equation}
where R$_{\rm acc}$ denotes the radius of the accretor. For the case of BH, R$\rm_{acc}$ is the radius of the inner most stable circular orbit, which we calculate with Equation (2.21) in \cite{bardeen...72}. 
Given the observationally inferred spin $a_* > 0.95$ \citep{gou...11}, we adopt the median $a_* = 0.97$ and find
\begin{equation}
R_{\rm acc} =  2.57 \left ( \frac{M_{\rm BH}}{M_\sun} \right ) \; \rm{km}.
\end{equation}
This calculated $\rm L_{bol}$ needs to fall within the observational range.

\begin{figure*}
\begin{center}
\includegraphics[clip=false]{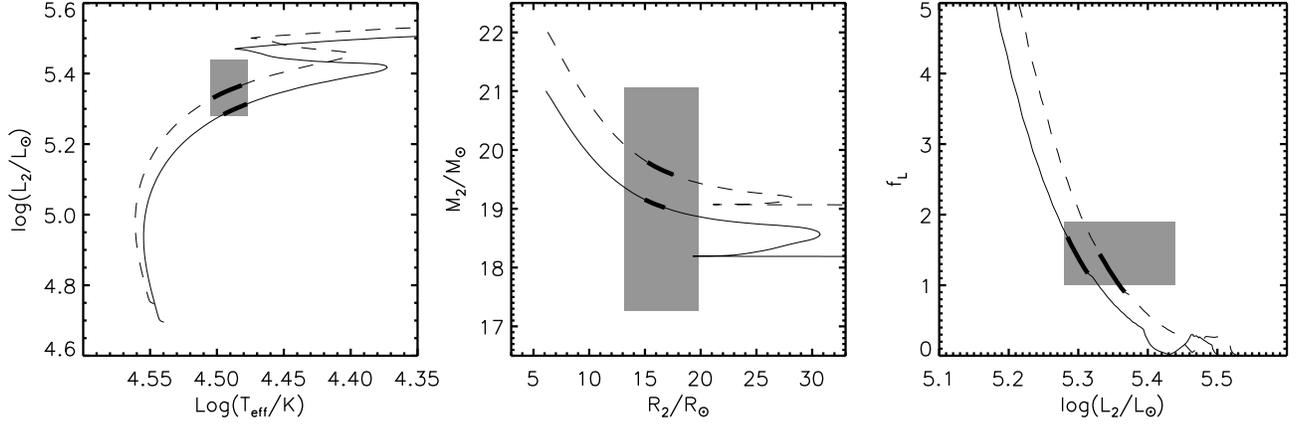}
\end{center}
\caption{Systemic behavior of two selected evolutionary sequences, which have the same $\alpha_{\rm ov} = 0.44$, M$_{\rm BH} = 14.81$ M$_\sun$, $\alpha_{\rm wind} = 1.5$, and $\beta = 1.0$. Sequence 1 (solid) and 2 (dashed) have M$\rm_{2,zams}$ of 21 and 22 M$_\sun$, respectively. The left panel shows the evolutionary tracks on the H-R diagram, while the middle panel illustrates the behaviors of the mass and the radius of the star. The right panel shows the variations of the calculated stellar luminosity and $f_{\rm L}$, where $f_{\rm L}$ is defined in Equation (10). The gray shaded areas represent the observational constraints on the relevant quantities, and the thick part of the evolutionary tracks indicates the part of the sequence that the observational constraints on the H-R diagram are satisfied.}
\end{figure*}

The last additional constraint is that the observational constraints on L$_2$ and L$\rm_{bol}$ have to be evaluated at the same distant estimation. To examine this, we calculate the ratio
\begin{equation}
f_L \ = \ \left ( \frac{L_2}{10^5 L_\sun} \right ) \left ( \frac{L_{bol}}{10^{37} erg s^{-1}} \right )^{-1} \ ,
\end{equation} 
which is independent of distance. From Figure\;1 in \cite{orosz...11}, L$_2$ is $2.09 \times 10^5$ L$_\sun$ at T$_{\rm eff} = 30000$K, and is $2.51 \times 10^5$ L$_\sun$ at T$_{\rm eff} = 32000$K, assuming a distance of 1.86 kpc. Together with the measured range of L$\rm_{bol}$ rescaled at the same distance estimation, the upper and lower limits of $f\rm_L$ are 1.01 and 1.90, respectively. We can assure that both luminosity constraints are evaluated at the same distance estimation if $f\rm_L$ falls within that range.

In order to find the current evolutionary stage of the BH companion, we apply these constraints to a set of evolutionary sequences, which cover the parameter space of the companion's ZAMS mass (M$\rm_{2,zams}$), $\rm \alpha_{ov}$, M$\rm_{BH}$, $\rm \alpha_{wind}$, and $\beta$. For each sequence, we find whether there exists a point in time that the calculated properties simultaneously satisfy all observational constraints: the BH companion's mass, luminosity, temperature, and radius, L$\rm_{bol}$, $f_{\rm L}$, and not overfilling the Roche lobe of the BH companion. Similar to the Roche lobe constraint, we also consider an uncertainty of $\pm{2.5}$ R$_\sun$ in the calculated stellar radii when we apply the observational constraint of the BH companion's radius. If such a period of time exists, we classify that evolutionary sequence as "successful". The behavior of some relevant parameters is illustrated in Figure\;2 for two selected successful sequences that are chosen mainly to provide a clear and instructive picture. The displayed sequences therefore do \em not \em represent our best possible matches to the observed properties of Cygnus X-1. Figure\;3 shows the parameter space of M$\rm_{2,zams}$, $\rm \alpha_{ov}$, and M$\rm_{BH}$ covered by \em all \em successful sequences. For $\rm \alpha_{wind}$ and $\beta$, the successful sequences covered the entire allowed parameter space, which are $\rm 1.5 \leq \alpha_{wind} \leq 2.0$ and $0.6 \leq \beta \leq 1.6$.

\begin{figure}
\begin{center}
\plotone{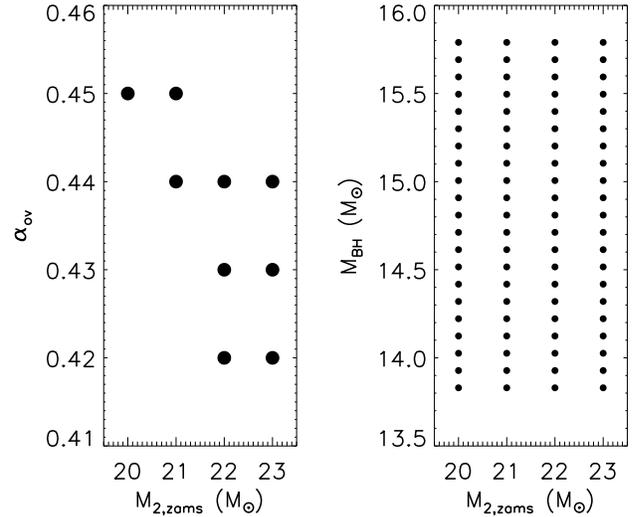}
\end{center}
\caption{The parameter space of M$\rm_{2,zams}$, $\rm \alpha_{ov}$, and M$_{\rm BH}$ covered by all successful sequences.}
\end{figure}

The current age of the BH companion could be derived from the time interval at which all observational constraints are satisfied. Assuming that the BH progenitor and its companion formed at the same time, we could compute the time since the BH formation (t$_{\rm sys}$) by 
\begin{equation}
t_{\rm sys} \; = \; t_2 \; - \; t_{\rm BH} \; ,
\end{equation}
where t$_{\rm BH}$ is the approximate lifetime of the BH progenitor. We follow \cite{belczynski...10} to calculate $\rm M_{BH}$ and $\rm t_{BH}$ for different progenitors using the stellar evolution code SSE \citep{hurley...00}, and adopting the mass loss prescriptions which were classified as "Vink et al. Winds". The calculated t$_{\rm BH}$ are fit as a function of M$_{\rm BH}$,
\begin{equation}
\frac{t_{\rm BH}}{10^6 \rm yrs} = \frac{\frac{M_{\rm BH}}{M_\sun}}{19.26 - 4.902 \left (\frac{M_{\rm BH}}{M_\sun} \right) + 0.3841 \left (\frac{M_{\rm BH}}{M_\sun} \right )^2} + 3.341,
\end{equation}
for $\rm M_{\rm BH} \geq 9.5$ M$_\sun$. Figure\;4 shows the variations of t$_2$ and t$_{\rm BH}$ against M$\rm_{2,zams}$. We find that t$\rm_{sys}$ is between 4.8 and 7.6 Myr.

\begin{figure}
\begin{center}
\plotone{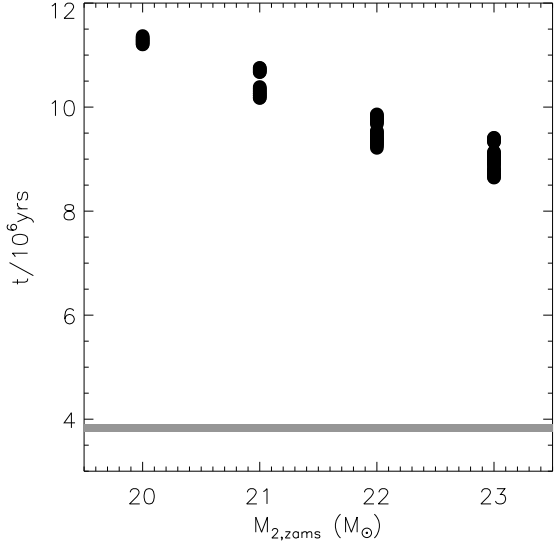}
\end{center}
\caption{The variations of t$_2$ (circles) against M$\rm_{2,zams}$. The gray shaded region indicates the range of t$\rm_{BH}$ for the corresponding successful sequences, which is calculated by Equation (12). The difference between t$_2$ and t$_{\rm BH}$ gives t$_{\rm sys}$.}
\end{figure}

\pagebreak
\section{Kinematic History in the Galaxy}

Here, we assume that Cygnus\;X-1 formed in the Galactic disk. The consideration of Cyg OB3 being the parent association of Cygnus\;X-1 is discussed in Section 9.2. Given the observed position and measured proper motion of Cygnus\;X-1, we derive the post-SN peculiar velocity of the binary's center-of-mass by tracing its orbit in the Galaxy back to the time of BH formation. We describe the motion of the binary with respect to a right-hand Cartesian reference frame, whose origin coincides with the Galactic center. The Z axis points to the northern Galactic pole, while the X axis points in the direction from the projected position of the Sun onto the Galactic plane to the Galactic center. In this reference frame, the Sun is located at (X$_\sun$,\;Y$_\sun$,\;Z$_\sun$) = ($-8.5,0,0.03$) kpc \citep{joshi07, ghez...08, gillessen...09, reid...09}, and has a peculiar motion (U$_\sun$,\;V$_\sun$,\;W$_\sun$) = (11.1, 12.24, 7.25) km\;s$^{-1} $\citep{schonrich...10}. Cygnus\;X-1 is currently located at a distance of $1.86^{+0.12}_{-0.11}$ kpc from the Sun, with a Galactic longitude $l$ = 71.3$^\circ$, and a Galactic latitude $b$ = 3.1$^\circ$ \citep{lestrade...99, reid...11}. This means Cygnus\;X-1 is currently $\sim 130$ kpc above the Galactic plane.

To model the Galaxy, we adopt the Galactic potential of \cite{carlberg-innanen87} with updated model parameters of \cite{kuijken-gilmore89}. The equations governing the system's motion in the Galaxy are integrated backward in time, up to the time corresponding to the current system's age t$_{\rm sys}$ given by the successful sequences. We follow the methodology of \cite{gualandris...05} to initialize the parameters for the integration, which accounts for the uncertainties in the estimated distance and measured velocity components. We generate the initial system's position by the Galactic coordinates ($l$,\,$b$) and a random distance drawing from a Gaussian distribution. We generate initial system's velocity by drawing randomly the proper motions ($\mu_{\rm R.A.}$,\,$\mu_{\rm decl.}$) and heliocentric radial velocity (V$_0$) from Gaussian distributions. The current system's age is uniformly distributed between 4.8 and 7.6 Myr (see Figure\;4).

\begin{figure}
\begin{center}
\plotone{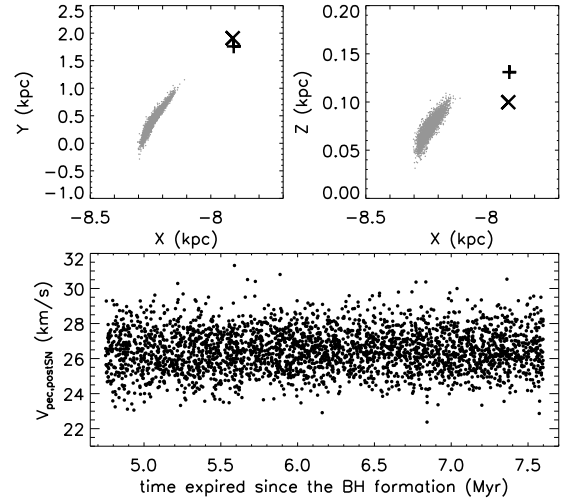}
\end{center}
\caption{Upper panels: The grey dots illustrate the possible locations of Cygnus X-1 at the birth time of the BH, obtained from 3,000 integrations of its trajectory backwards in time. The initial conditions of the integrations are generated randomly using the methodology described in Section\;5. The plus signs indicate the current location of Cygnus X-1, derived from the mean distance of 1.86 kpc. The crosses represent the current location of Cyg OB3 center, with an adopted distance of 2 kpc. Lower panel: The distribution of post-SN peculiar velocities V$_{\rm pec,postSN}$ against the time expired since the BH formation.}
\end{figure}
Figure\;5 shows the possible positions of Cygnus\;X-1 at the time of BH formation, obtained from integrating 3,000 trajectories backwards in time. As there is no trajectory crossing the Galactic plane and the end points of all trajectories fall within 110\;pc from the Galactic plane, we consider each end point as a possible birth site of the BH. The post-SN peculiar velocity V$_{\rm pec,postSN}$ of the binary is obtained by subtracting the local Galactic rotational velocity from the center-of-mass velocity of the binary at the birth sites. We find V$_{\rm pec,postSN}$ ranges from 22 to 32 km\;s$^{-1}$ and is time independent. The distribution of V$_{\rm pec,postSN}$ against the time expired since the formation of the BH are displayed in Figure\;5.

\section{Orbital Dynamics at Core Collapse}

For each of the successful sequence, we perform a Monte Carlo simulation which consists of twenty million pre-SN binaries. The properties of the BH progenitor's companion are taken from the stellar model of that sequence, at the time when the age of the star is equal to t$_{\rm BH}$. During a supernova (SN) explosion, the mass loss from the system and possibly the kick imparted to the BH change the binary's orbital parameters. The pre- and post-SN component masses, orbital semi-major axis, and orbital eccentricity are related by the conservation laws of the orbital energy and angular momentum. In the followings, we add the subscripts "preSN" and "postSN" to the notations of the orbital elements to distinguish between their values just prior and right after the SN explosion that formed the BH. 

We start with seven free parameters: the BH immediate (He-rich) progenitor mass (M$_{\rm He}$), pre-SN orbital semi-major axis (A$_{\rm preSN}$) and eccentricity (e$_{\rm preSN}$), the mean anomaly ($m$), the magnitude (V$\rm_k$) and direction ($\theta$, $\phi$) of the kick velocity imparted to the BH. $\theta$ is the polar angle of the kick with respect to the relative orbital velocity of the BH progenitor just prior to the SN explosion, and $\phi$ is the corresponding azimuthal angle \citep[see Figure 1 in][for a graphical representation]{kalogera00}. The first five parameters are drawn from uniform distributions, while the last two are drawn from isotropic distributions. It is obvious that the progenitor must of course be more massive than the BH, but there is no absolute upper limit for the progenitor mass. We adopt M$_{\rm He} \le$20 M$_\sun$, and provide a discussion on this upper limit in Section 9.1.

The relations between pre- and post-SN parameters have been derived by \cite{hills83}:
\begin{align}
V^2_k \; + \; V^2_{\rm He,preSN}& \; + \; 2V_kV_{\rm He,preSN}\cos \theta \nonumber \\
	 = \; G(&M_{\rm BH}+M_2)\left (\frac{2}{r} - \frac{1}{A_{\rm postSN}} \right ), \\
G (M_{\rm BH} + M_2) & A_{\rm postSN}(1-e^2_{\rm postSN}) \nonumber \\
	= r^2 \Bigl ( & V^2_k \sin^2 \theta \cos^2 \phi + \bigl [ \sin \psi (V_{\rm He,preSN} + V_k \cos \theta) \nonumber \\
	& - V_k \cos \psi \sin \theta \sin \phi \bigr ]^2 \Bigr ) ,
\end{align}
Here, r is the orbital separation between the BH progenitor and its companion at the time of SN explosion, 
\begin{equation}
r \;=\; A_{\rm preSN}(1\:-\:e_{\rm preSN}\cos\:E_{\rm preSN}),
\end{equation}
where $E$ is the eccentric anomaly, and is related to $m$ as
\begin{equation}
m\;=\;E\;-\;e\: \sin E.
\end{equation}
V$_{\rm He,preSN}$ is the relative pre-SN orbital velocity of the BH progenitor,
\begin{equation}
V_{\rm He,preSN} = \left [ G(M_{\rm He}+M_2)\Bigl ( \frac{2}{r} - \frac{1}{A_{\rm preSN}}\Bigr ) \right ]^{1/2}.
\end{equation}
The angle $\psi$ is the polar angle of the position vector of the BH with respect to its pre-SN orbital velocity in the companion's frame. It is related to the pre-SN parameters as
\begin{equation}
r^2V^2_{\rm He,preSN} \sin^2 \psi \; = \; G(M_{\rm He} + M_2) A_{\rm preSN} (1 - e^2_{\rm preSN}).
\end{equation}
Since the core collapse is instantaneous, r remains unchanged. This gives a constraint
\begin{align}
r \; & =\; A_{\rm preSN}(1\:-\:e_{\rm preSN}\cos E_{\rm preSN}) \nonumber \\
	& =\; A_{\rm postSN}(1\:-\:e_{\rm postSN}\cos E_{\rm postSN}),
\end{align}
which needs to be satisfied with $|\cos E_{\rm postSN}| \leqslant 1$. 

The mass loss from the system and a natal kick imparted to the BH can induce a post-SN peculiar velocity (V$_{\rm pec,postSN}$) at the binary's center of mass. Its magnitude is determined by  following Equations (28)--(32) in Paper I, and is required to fall within the range derived in Section 5, which is $22 - 32$ km/s.

In addition, there are two more restrictions on the properties of pre- and post-SN binary components. First, we require that both components have to fit within their pre- and post-SN Roche lobe at periapsis. We impose this condition to avoid complications arising from mass transfer induced changes in the stellar structure of the MS companion, that later becomes the BH companion of the XRB. To calculate the Roche lobe radius of each component in eccentric pre- and post-SN orbits, we adopt the fitting formulae of \cite{sepinsky...07}. When calculating the pre-SN Roche lobe radii, we assume that the pre-SN orbit is pseudo-synchronized. Again, due to the difference in calculated stellar radii among stellar evolution codes, we consider an uncertainty of $\pm{2.5}$ R$_\sun$ on the companion radius \citep{valsecchi...10}. The radius of the BH immediate progenitor can be approximated by Equations (3) in \cite{fryer-kalogera97}, since we assume that it is a Helium star. Second, the pre-SN spin of the BH immediate progenitor and its companion need to be less than the breakup angular velocity $\rm \Omega_c$ $\approx$ $\rm (GM/R^3)^{1/2}$. As the calculated stellar radius R$_2$ associates with an uncertainty $\Delta \rm R = 2.5 R_\sun$ \citep{valsecchi...10},
\begin{equation}
\Omega_c = \sqrt{\frac{GM_2}{R_2^3} } \left ( 1 + \frac{3}{2} \frac{\Delta R}{R_2}\right )
\end{equation}
for the BH companion.

\section{Orbital Evolution After the SN Explosion}

The orbital evolution of the simulated binaries, which are generated from the Monte Carlo simulations described in Section\;6, is calculated up to the current epoch. After the formation of the BH, the orbital parameters of the binary are subject to secular changes due to the tidal torque exerted by the BH on its companion, and due to the loss of orbital angular momentum via gravitational radiation and stellar wind. Since the tidal interactions depend on both the orbital and rotational properties of the MS companion, the star's rotational angular velocity ($\Omega$) right after SN explosion that formed the BH enters the problem as an additional unknown quantity. Here we assume the rotational angular velocity of the BH companion is unaffected by the SN explosion, and is pseudo-synchronized to the pre-SN orbital frequency. The system of equations governing the tidal evolution of the orbital semi-major axis A, eccentricity $e$, and the BH companion's rotational angular velocity $\Omega$ has been derived by \cite{hut81}:
\begin{align}
\left( \frac{dA}{dt} \right )_{\rm tides}  = \; & -\, 6 \frac{k_2}{T} \frac{M_{\rm BH}}{M_2} \frac{M_{\rm BH} + M_2}{M_2} \left ( \frac{R_2}{A}\right )^8 \nonumber \\
	& \times \frac{A}{(1-e^2)^{15/2}} \left [ f_1\left(e^2\right) - \left(1-e^2\right)^{3/2}f_2\left(e^2\right) \frac{\Omega}{n}\right ], \\
\left( \frac{de}{dt} \right )_{\rm tides}  = \; & -\, 27 \frac{k_2}{T} \frac{M_{\rm BH}}{M_2} \frac{M_{\rm BH} + M_2}{M_2} \left ( \frac{R_2}{A}\right )^8 \nonumber \\
	& \times \frac{e}{(1-e^2)^{13/2}} \left [ f_3\left(e^2\right) - \frac{11}{18}\left(1-e^2\right)^{3/2}f_4\left(e^2\right) \frac{\Omega}{n}\right ], \\
\left( \frac{d\Omega}{dt} \right )_{\rm tides}  = \; & 3 \frac{k_2}{T} \left ( \frac{M_{\rm BH}}{M_2} \right )^2 \frac{M_2R_2^2}{I_2} \left ( \frac{R_2}{A} \right )^6 \nonumber \\
	& \times \frac{n}{(1-e^2)^6} \left [ f_2\bigl(e^2\bigr) - \bigl(1-e^2 \bigr)^{3/2}f_5 \bigl (e^2 \bigr ) \frac{\Omega}{n} \right ].
\end{align}
Here, $k_2$ and $I_2$ are the apsidal-motion constant and moment of inertia of the MS companion, respectively. T is a characteristics timescale for the orbital evolution due to tides, and n = 2$\pi$/P$_{\rm orb}$ is the mean orbital angular velocity. The coefficient functions $f_i\left(e^2\right)$ for i = i, 2, $\dotso$, 5 are given in Equations (11) in \cite{hut81}. As the BH companion in Cygnus\;X-1 is a massive MS star with a radiative envelope, the factor $k_2/T$ can be approximated as
\begin{align}
\left ( \frac{k_2}{T} \right )_{\rm rad} = \; &1.9782 \times 10^4 \left ( \frac{R_2}{R_\sun} \right ) \left ( \frac{R_\sun}{A} \right )^{5/2} \nonumber \\
	& \times \left ( \frac{M_2}{M_\sun} \right )^{1/2} \left (\frac{M_{\rm BH}+M_2}{M_2} \right )^{5/6} E_2 \; \; yr^{-1}.
\end{align}
The constant E$_2$ comes from a fit to the tables in \cite{claret04},
\begin{equation}
log E_2 = - \frac{t / t_{ms}}{2.20489 - 1.89579 (t / t_{ms})} - 5.51039,
\end{equation}
for 15.85 $\rm \leq M_{2,zams} \leq 25.12$ M$_\sun$. Here, t$\rm_{ms}$ is the main sequence lifetime. We define the end of the main sequence as the hydrogen abundance at the core being less than 0.01.

To follow the secular changes of the orbital parameters associated with emissions of gravitational waves, we adopt Equations (35) and (36) in \cite{junker-schaefer92}, which are derived up to 3.5 post-Newtonian order.

The rates of change in A and e due to wind mass loss and wind accretion onto the BH are determined by following Equations (15) and (16) in \citep{hurley...02},
\begin{align}
\left ( \frac{dA}{dt} \right )_{\rm wind} = &\; -A \left [ \frac{\dot{M}_2}{M_{\rm BH}+M_2} + \left ( \frac{2-e^2}{M_{\rm BH}} + \frac{1+e^2}{M_{\rm BH}+M_2}\right ) \frac{\dot{M}_{\rm acc}}{1-e^2} \right ] \, \\
\left ( \frac{de}{dt} \right )_{\rm wind} = & \; -e\dot{M}_{\rm acc} \left ( \frac{1}{M_{\rm BH}+M_2} + \frac{1}{2M_{\rm BH}}\right )\, .
\end{align}

The mass loss via stellar wind also induces a loss in the spin angular momentum of the BH companion. \cite{hurley...00} showed that if all the mass is lost uniformly from a thin shell at the surface of the MS star,
\begin{equation}
\dot{J}_{\rm 2,spin} \; = \frac{d}{dt} \left ( I_2 \Omega \right ) \; = \; \frac{2}{3}\dot{M}_2 R^2_2 \Omega \; ,
\end{equation}
where $J_{\rm 2,spin}$ is the spin angular momentum of the BH companion.

\begin{figure}
\begin{center}
\plotone{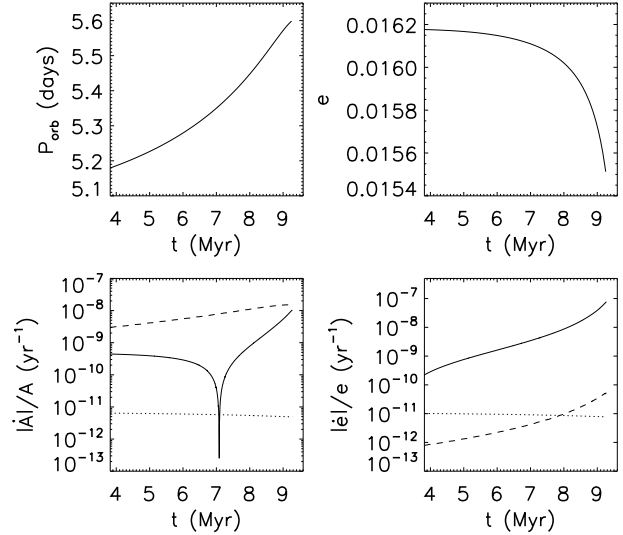}
\end{center}
\caption{The orbital evolution of a selected winning binary. Right after the formation of the BH (t = 3.8 Myr), this binary consists of a 14.8 M$_\sun$ BH and a 21.7 M$_\sun$ main sequence star. The top panels show the time evolution of orbital period and eccentricity. The bottom panels show the rate of changes of the semi-major axis A and eccentricity e due to tidal effects (solid line), wind mass loss and wind accretion onto the BH (dashed line), and gravitational radiation (dotted line).}
\end{figure}

For each of simulated binaries, we follow the secular changes of its orbital properties due to \em all \em the mechanisms mentioned in this section. The properties of binary components are adopted from the corresponding successful sequence. Unlike finding V$\rm_{pec,postSN}$ in Section\;5, the orbital evolution of the binary goes forward in time,  from $t_{\rm BH}$ to $t_2$. Within this period of time, the BH companion has to always fit within its Roche lobe at periapsis. In other words, its calculated radius is constrained to be less than the Roche lobe radius at periapsis given by \cite{sepinsky...07}. Again, we allow an uncertainty of $\pm{2.5}$ R$_\sun$, due to the difference in calculated stellar radii among stellar evolution codes \citep{valsecchi...10}. Furthermore, the rotational angular velocity of the BH companion has to be smaller than the breakup angular velocity $\Omega_c$. If the orbital period and eccentricity of the simulated binary at $t_2$ match the measured values of Cygnus\;X-1, we classify that binary as a "winning binary". Figure\;6 illustrates the time evolution of orbital parameters for one selected winning binary, and shows that the change in the semi-major axis is mainly determined by the stellar wind mass loss, while the change in the eccentricity is overwhelmingly dominated by the tidal effects.

\section{Progenitor Constraints}

The elements presented in the previous sessions can now be combined to establish a complete picture of the evolution of Cygnus X-1 and the dynamics involved in the core collapse event that formed the BH. After finding the successful evolutionary sequences that satisfy all the observed properties of the BH companion and the bolometric X-ray luminosity as discussed in Section\;4, we trace the motion of the system in the Galaxy back in time to the formation of the BH. We adopt the methodology of \cite{gualandris...05} to account for the uncertainties in the measured distance and velocity components of Cygnus\;X-1. The time of BH formation is different for each successful sequence. It is estimated by the BH mass of the sequence, which connects to an approximate lifetime of the corresponding BH progenitor. This procedure gives us a constraint on the system's peculiar velocity right after the BH formation. We then perform Monte Carlo simulations on the orbital dynamics at core collapse for \em each \em successful sequence. There are seven free parameters: the BH immediate progenitor mass, the pre-SN orbital semi-major axis and eccentricity, the mean anomaly, the magnitude of kick velocity imparted to the BH, and the two angles specifying the direction of the kick velocity. The Monte Carlo simulations produce a population of simulated binaries, which satisfy the post-SN system's peculiar velocity constraint derived already. Last, we evolve the orbits of these simulated binaries forward in time to the current epoch. If the orbital period and eccentricity of the simulated binary at current epoch match the measured values of Cygnus\;X-1, we classify that simulated binary as a "winning binary". The results presented in what follows are derived from the winning binaries of \em all \em successful sequences.

In Figure\;7, we present the probability distribution functions (PDFs) of the BH immediate (He-rich) progenitor mass (M$\rm_{He}$) and natal kick magnitude (V$\rm_k$). We find M$_{\rm He}$ to be in a range of $15.0 - 20.0$ M$_\sun$, and V$_{\rm k}$ to be $\le 77$ km\;s$^{-1}$, both at 95\% confidence. Figure\;8 illustrates the 2D joint V$_{\rm k}$--M$_{\rm He}$ confidence levels, which shows that if M$_{\rm He}$ is less than $\sim$17 M$_\sun$, the BH might have received a non-zero natal kick at the core collapse event. For small M$_{\rm He}$, a minimum V$_{\rm k}$ of  $\sim 55$ km s$^{-1}$ is \textit{necessary} for explaining the current observed properties of Cygnus\;X-1. Furthermore, both the M$\rm_{He}$ PDF and the 2D joint V$_{\rm k}$--M$_{\rm He}$ confidence levels show that the maximum M$\rm_{He}$ is constrained by our adopted upper limit of 20 M$_\sun$. We impose this limit based on the physics involved in the evolution of massive stars. A discussion on this limit can be found in Section\;9.1. Given our understanding of mass loss from Helium stars, it seems that the BH has potentially received a small natal kick velocity of $\le 77$ km s$^{-1}$ (95\% confidence) during the core collapse event.

\begin{figure}
\begin{center}
\plotone{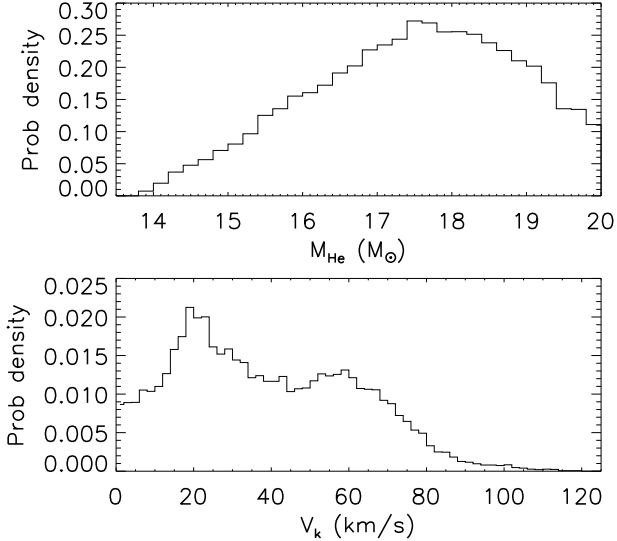}
\end{center}
\caption{The probability distribution functions of the BH immediate (He-rich) progenitor mass (M$_{\rm He}$) and natal kick magnitude (V$_{\rm k}$) imparted to the BH.}
\end{figure}

\begin{figure}
\begin{center}
\plotone{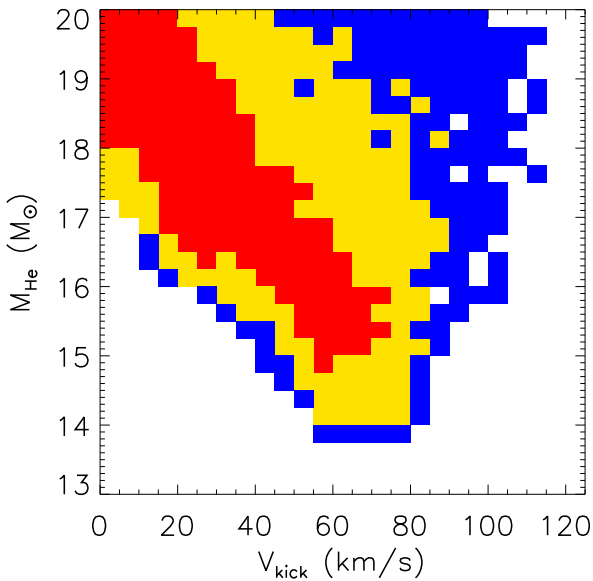}
\end{center}
\caption{The 2D joint V$_{\rm k}$--M$_{\rm He}$ confidence levels: 68.3\% (red), 95.4\% (yellow), and 99.7\% (blue).}
\end{figure}

Based on the dynamical model of \cite{orosz...11}, \cite{gou...11} found that the BH in Cygnus\;X-1 has a spin parameter $a_* > 0.95$ at 3$\sigma$. To determine whether the BH was born with an extreme spin, we first estimate how much mass the BH could have accreted from its companion's stellar wind since the time of BH formation. The winning binaries of all successful sequences show that at maximum the BH has accreted $\sim 2 \times 10^{-3}$ M$_\sun$. Since it is impossible to spin the BH up to $a_* > 0.95$ by accreting that negligible amount of mass, the BH needs to have an extreme spin at birth. This high spin has implications about BH formation and the role of rotation in core collapse. \cite{axelsson...11} also concluded that the observed spin connects to processes involved in core collapse, and is not likely to originate from the synchronous rotation of the BH progenitor.

Besides the constraints on the BH formation, our results also shed light on the evolutionary picture of Cygnus\;X-1. We find that right after the formation of the BH, the BH companion has a mass of $19.8 - 22.6$ M$_\sun$, in an orbit with period of $4.7 - 5.2$ days. Since then, the orbital separation of Cygnus\;X-1 has been increasing with time, as the rate of change in the semi-major axis is dominated by the influence of stellar wind mass loss from the system. On the other hand, the orbital eccentricity has decreased slightly since the BH formation. This is because the tides exerted on the companion by the BH, as the dominant mechanism of circularizing the orbit, are not strong enough to decrease the orbital eccentricity significantly within the time period of several million years since the time of BH formation. We find that e$\rm_{postSN}$ ranges from 0.015 to 0.022. However, this does not suggest that e$\rm_{preSN}$ has to be small. An eccentric pre-SN orbital could become fairly circular if there is a natal kick imparted to the BH at the right direction. As illustrated in Figure\;9, there are winning binaries with e$\rm_{preSN}$ being as high as $\sim 0.53$.

\begin{figure}
\begin{center}
\plotone{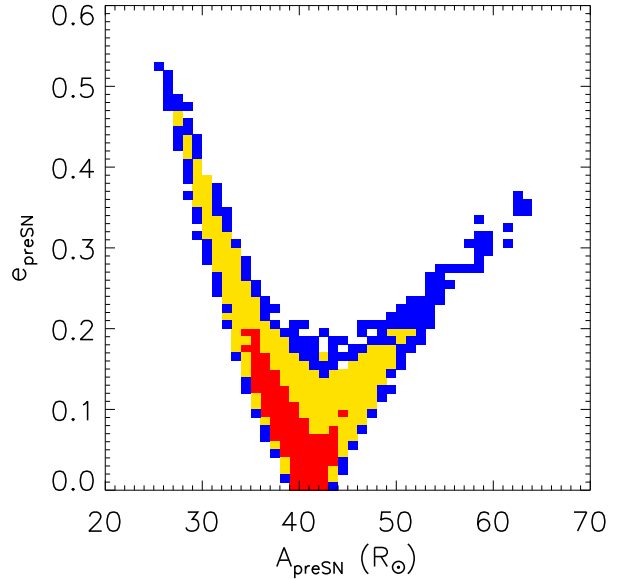}
\end{center}
\caption{The 2D joint A$_{\rm preSN}$--e$_{\rm preSN}$ confidence levels: 68.3\% (red), 95.4\% (yellow), and 99.7\% (blue).}
\end{figure}

\section{Summary \& Discussion}

In this paper we constrained the progenitor properties and the formation of the BH in the persistent XRB Cygnus\;X-1. Our analysis accounts for the orbital evolution and motion through the Galactic potential right after the BH formation, and the binary orbital dynamics at the time of core collapse. We find that the mass of the BH immediate progenitor falls within a range of $15.0 - 20.0$ M$_\sun$ at 95\% confidence. We note that the maximum progenitor mass is constrained by our adopted upper limit, which is discussed in Section\;9.1. The BH has potentially received a small natal kick velocity of $\le 77$ km s$^{-1}$ at 95\% confidence. In fact if the progenitor mass is less than $\sim 17$ M$_\sun$, a non zero natal kick velocity is \textit{necessary} to explain the currently observed properties of Cygnus\;X-1. Since the BH has only accreted mass from its companion's stellar wind, the total amount of mass accreted since the BH formation is less than $\sim 2 \times 10^{-3}$ M$_\sun$. This indicates that the observationally inferred BH spin of $a_* > 0.95$ \citep{gou...11} cannot be explained by mass accretion and has to be natal. This high spin has implications about BH formation and the role of rotation in core collapse. Right after the BH formation, the BH companion has a mass of $19.8 - 22.6$ M$_\sun$, in an orbit with period of $4.7 - 5.2$ days and eccentricity of $0.015 - 0.022$. Although the post-SN orbital eccentricity is small, the pre-SN orbit can potentially be fairly eccentric. This is possible if the BH receives a natal kick velocity at the right magnitude and direction. 

The formation of the BH in Cygnus\;X-1 has been previously studied by \cite{nelemans...99} and \cite{Mirabel-Rodrigues03}. Both studies assumed symmetric mass loss during the core collapse event, and considered only the binary orbital dynamics at the time of core collapse. Comparing with these two earlier studies, we consider the possible asymmetries developed during the core collapse event and the evolution of the binary since the BH formation. It is important to note that these two earlier studies do not consider the multitude of the observational constraints taken into account here and hence the suggested progenitors are not complete solutions for the evolutionary history of Cygnus\;X-1.

Finally, we discuss some of the assumptions introduced in our analysis in the following sub-sections.

\subsection{Maximum BH Progenitor Mass}

Unlike the case of GRO J1655-40 studied in Paper I, the analysis of orbital dynamics during the core collapse event does not give an upper limit on M$_{\rm He}$. Instead, we have conservatively adopted an upper limit of M$_{\rm He}$ $\le$20 M$_\sun$, based on physics involved in the evolution of massive stars. As mentioned in Section 6.1 of Paper II, by evolving a ZAMS star of $\sim$100 M$_\sun$ at solar metallicity, the maximum Helium star mass one can achieve is $\sim$15 M$_\sun$ when including moderate stellar rotation, and $\sim$17.5 M$_\sun$ when assuming no stellar rotation. When adopting the upper limit of 17.5 M$_\sun$, the lower limit of M$\rm_{He}$ decreases slightly to 14.6 M$_\sun$ and the range of V$\rm_{k}$ becomes $14 - 81$ km\;s$^{-1}$, both with 95\% confidence. This range of V$\rm_k$ still suggests that the BH in Cygnus\;X-1 received a low kick during the core collapse event.

\subsection{Association with Cyg OB3}

The center of Cyg OB3 locates at $l = 72.8^\circ$ and $b = 2.0^\circ$, and at a distance of $1.4 - 2.7$ kpc away from the Sun \citep{massey...95, dambis...01, melnik...01, melnik-dambis09}. When comparing that to the location of Cygnus\;X-1 (Table\;1), it is clear that not only their Galactic coordinates are close to each other, but also their distance estimations overlap with each other. Furthermore, the measurements of proper motion and radial velocity show that Cygnus X-1 is moving as the members of Cyg OB3 \citep{dambis...01, Mirabel-Rodrigues03, melnik-dambis09}. Based on these observations, \cite{Mirabel-Rodrigues03} argue that Cyg OB3 is the parent association of Cygnus X-1. This infers that V$_{\rm pec,postSN}$ due to the core collapse event has to be small. If we change the constraint on V$_{\rm pec,postSN}$ to $\le 10$ km s$^{-1}$, we find M$_{\rm He}$ to be in a range of $13.9 - 16.9$ M$_\sun$ and V$_{\rm k}$ to be $\le 24$ km/s, both at 95\% confidence. Besides the change in 95\% limits, non-zero BH natal kicks are not needed for progenitors of M$_{\rm He} \le 17$ M$_\sun$ in order to explain the observed properties of Cygnus\;X-1, but become necessary for M$_{\rm He} > 17.5$ M$_\sun$ (see Figure\;10). Also, we note that a relatively small change on the range of V$\rm_{pec,postSN}$ affects the derived constraint on V$_{\rm k}$ qualitatively.

\begin{figure}
\begin{center}
\plotone{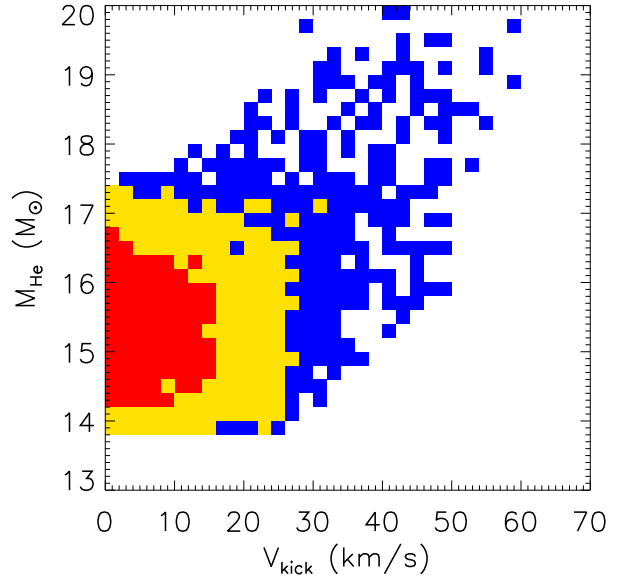}
\end{center}
\caption{The same plot of 2D joint V$_{\rm k}$--M$_{\rm He}$ confidence levels as Figure\;8, calculated with V$\rm_{pec,postSN} \le 10$ km s$^{-1}$ instead of the original range derived Section\;5.}
\end{figure}

\subsection{Super-Synchronized Orbit}

After considering several previous measurements of the BH companion's surface rotation speed (V$_{\rm rot} \sin i$), \cite{caballero...09} adopted V$_{\rm rot} \sin i = 95 \pm{6}$ km s$^{-1}$. \cite{orosz...11} found that the ratio of the BH companion's spinning frequency to the orbital frequency ($f_{\Omega}$) was $1.400 \pm{0.084}$, which was derived based on their results of the inclination angle $i = 27^\circ.06 \pm{0^\circ.76}$ and the companion radius R$_2 = 16.5 \pm{0.84}$. This indicates that the BH companion is super-synchronized. We note that with the analysis presented here, we find none of our winning binaries have super-synchronized BH companions at the current epoch. They are all sub-synchronized with $f_{\Omega}$ reaching $\sim 0.87$ at maximum.

In an effort to examine how our standard assumptions can be modified and investigate whether super-synchronism is at all allowed by the models as indicated by the observations, we make two modifications to our analysis. We first remove the assumption that the pre-SN orbit is pseudo-synchronized, and randomly distribute the pre-SN spin of the BH companion between zero and its breakup angular frequency $\rm \Omega_c$. Next, we reduce the secular changes of the orbital parameters due to the influence of tides by multiplying the right hand side of Equation (21) -- (23) by a constant $f\rm_{tide}$. 

\begin{figure}
\begin{center}
\plotone{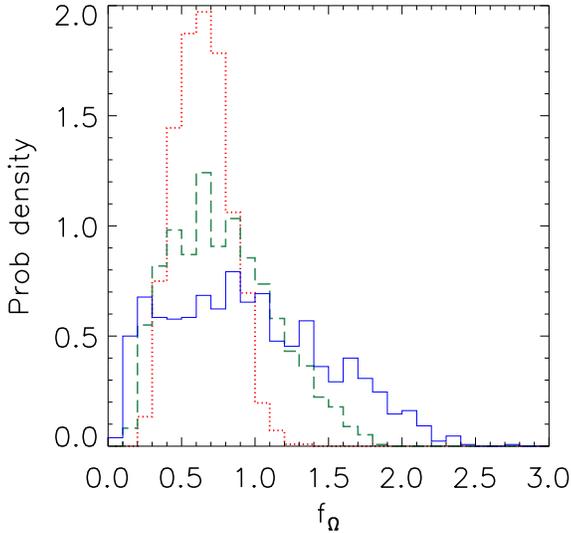}
\end{center}
\caption{The probability distribution functions of the ratio ($f_{\Omega}$) of the companion's spinning frequency to the orbital frequency at current epoch, which are calculated with models of $f\rm_{tide}$ = 0.2 (solid/blue), 0.5 (dashed/green), and 1.0 (dotted/red). All three set of models have the pre-SN spin of the companion randomly distributed between zero and the breakup angular frequency $\Omega_{\rm c}$.}
\end{figure}

As shown in Figure\;11, by allowing the pre-SN spin of companion to be greater than pseudo-synchronization and keeping the tidal strength unchanged (i.e.  $f\rm_{tide} = 1.0$), the maximum $f_{\Omega}$ of the winning binaries increases to $\sim 1.2$. Although it is getting close, this value is still below the observationally inferred one. Together with a weakened tidal strength of $f\rm_{tide} = 0.2$ and 0.5, we comfortably find winning binaries with $f_{\Omega} = 1.4$. Furthermore, the minimum pre-SN surface rotation speed of the companion in those winning binaries are $\sim 500$ and 700 km s$^{-1}$ for $f\rm_{tide} = 0.2$ and 0.5, respectively. Given the uncertainties in the physics of fast rotating massive stars, it seems that super-synchronism in Cygnus\;X-1 is allowed by our models presented, if the companion is spinning faster than orbital pseudo-synchronization right before the core collapse event and the tides exerted on the companion are weaker than the nominal theoretical values.  

\subsection{Constraints on the BH Kick Direction}

Since the measured eccentricity of Cygnus\;X-1 is very low (see Table\;1), and the orbit has not been circularized much since the formation of the BH (see Figure\;6), e$\rm_{postSN}$ has to be very small too. As discussed in Section\;8, we indeed find that e$\rm_{postSN}$ falls within a range of $0.015 - 0.022$. From the orbital dynamics at core collapse, the very low e$\rm_{postSN}$ might shed light on the properties of the natal kick imparted to the BH.

During the core collapse event, e$\rm_{postSN}$ is affected by the amount of mass loss, the direction and magnitude of the natal kick, as well as e$\rm_{preSN}$. As mass loss tends to increase the eccentricity, a natal kick in the right direction and magnitude is needed in order to counteract the effect of mass loss and result in a very low e$\rm_{postSN}$. It turns out that with the observationally inferred constraints on M$\rm_{BH}$, M$_2$, V$\rm_{pec,postSN}$, and the extra constraints mentioned in Section\;6, the eccentricities right after symmetric explosions (i.e. no natal kicks) are typically $\sim0.1$, and overwhelmingly $< 0.45$. Also, the difference between pre- and post-SN eccentricities is mostly $< 0.15$. These link to the requirement of the small V$\rm_{pec,postSN}$. Larger amounts of mass loss relative to the total mass of the pre-SN system not only leads to higher e$\rm_{postSN}$, but also larger V$\rm_{pec,postSN}$, and hence they are not allowed. For the same reason, the required kick velocity for getting the very low post-SN eccentricity also needs to be small. In addition, we note that under special conditions, e$\rm_{postSN}$ could be low even though e$\rm_{preSN}$ is high. \cite{hills83} showed that if the symmetric supernova explosion occurs at the proximity of apastron, e$\rm_{postSN}$ could be $\sim 0$ for a specific range of mass loss.

As the eccentricity induced by mass loss is low, natal kicks do not have to contribute dramatically to make e$\rm_{postSN}$ fall within the observationally required range. Although natal kicks have to be constrained in some directions, that constraint is not super narrow. By extracting data from the winning binaries, we found that 70\% have $\cos(\theta) < 0$; the distribution in the azimuthal angle $\phi$ for the kicks of the winning binaries does not deviate much from the a priori flat distribution. As a result, the requirement of a very low e$\rm_{postSN}$ does constrain the natal kick directions, but the constraint is not particularly strong.

\acknowledgments
This work has been primarily supported by the NSF Grant AST-0908930; VK also acknowledges partial support through the NSF Grant PHY-1066293 and the hospitality of the Aspen Center for Physics; TF acknowledges Fellowship support by the Harvard-Smithsonian Center for Astrophysics and the Harvard Institute for Theory and Computation.

\end{document}